\newcommand{\be}{\begin{equation}}
\newcommand{\ee}{\end{equation}}
\newcommand{\bea}{\begin{eqnarray}}
\newcommand{\eea}{\end{eqnarray}}
\newcommand{\nn}{\nonumber\\}
\newcommand{\p}[1]{(\ref{#1})}

\newcommand{\Db}{{\bar D}}
\newcommand{\tb}{{\bar\theta}}

\newcommand{\bF}{{\bar F}}

\newcommand{\vfi}{{\varphi}}
\newcommand{\bvfi}{{\bar\varphi}}
\newcommand{\ve}{\varepsilon}
\documentstyle[12pt]{article}
\begin{document}
\renewcommand{\thefootnote}{\fnsymbol{footnote}}
\begin{titlepage}
\mbox{} \vskip 2.6truecm
\begin{center}
{\Large\bf
 Goldstone Superfield Actions for Partially Broken \break AdS$_{\bf 5}$ Supersymmetry}
\end{center}
 \vskip 0.6truecm
 \begin{center}
{\large\bf S. Bellucci}\footnote{bellucci@lnf.infn.it } \vspace{0.5cm} \\
{\it INFN-Laboratori Nazionali di Frascati, \\
 C.P. 13, 00044 Frascati, Italy \vspace{0.5cm} }\\
{\large\bf E. Ivanov\footnote{eivanov@thsun1.jinr.ru }
and S. Krivonos}\footnote{krivonos@thsun1.jinr.ru }
  \vspace{0.5cm} \\
{\it Bogoliubov  Laboratory of Theoretical Physics, JINR,\\
141980 Dubna, Moscow Region, Russia} \vspace{1.5cm}
\end{center}
\vskip 0.6truecm  \nopagebreak
\begin{abstract}
\noindent We explicitly construct $N=1$ worldvolume supersymmetric
minimal off-shell Goldstone superfield actions for two options of $1/2$ partial
spontaneous breaking of AdS$_5$ supersymmetry $SU(2,2|1)$ corresponding
to its nonlinear realizations in the supercosets with
the AdS$_5$ and AdS$_5\times S^1$ bosonic parts. The relevant
Goldstone supermultiplets are comprised, respectively, by improved tensor and
chiral $N=1$ superfields. The second action is obtained
from the first one by duality transformation. In the bosonic sectors they
yield static-gauge Nambu-Goto actions for L3-brane on AdS$_5$ and scalar 3-brane
on AdS$_5\times S^1$.

\end{abstract}
\newpage

\end{titlepage}

\renewcommand{\thefootnote}{\arabic{footnote}}
\setcounter{footnote}0
\setcounter{equation}0
\noindent{\bf 1. Introduction.} The concept of partial breaking of global
supersymmetry (PBGS) \cite{bw}
provides a manifestly worldvolume supersymmetric description of various
superbranes in terms of Goldstone superfields \cite{pbgs}.

Most of the PBGS theories known to date correspond to superbranes
on flat super Minkowski backgrounds (see \cite{bik0,i1} and refs. therein).
On the other hand, keeping in mind the renowned AdS/CFT correspondence \cite{ads},
it is the AdS$_n\times S^m$ and PP-wave type \cite{mets} superbackgrounds
which are of primary interest.
However, not too many explicit examples of the worldvolume superfield PBGS actions
on such backgrounds were constructed so far. Such actions were given only for
$N=1$ supermembrane in AdS$_4$ \cite{dik} and some its dimensional reductions
\cite{ik,ikn}.

It is tempting to construct PBGS versions of superstring and D3-brane on
the AdS$_5\times S^5$ background which is in the heart
of the original AdS/CFT conjecture. These systems
should be associated with the partial breaking of $N=4, d=4$ superconformal
group $SU(2,2|4)$ which determines the corresponding superisometries.
\footnote{The space-time Green-Schwarz-type actions for
these systems were constructed in \cite{mt1,mt2}.}
It is natural to firstly study some truncations of these models
based on simpler $N=1$ and $N=2$, $d=4$ superconformal groups
$SU(2,2|1)$ and $SU(2,2|2)$. An attempt to construct a PBGS model for $SU(2,2|1)$ which
would generalize that of \cite{BG2}
was undertaken in \cite{kum}. This model involves Goldstone $N=1$ chiral
superfield as the basic one and is expected
to describe a scalar 3-brane on AdS$_5\times S^1$. However, no proper
Goldstone superfield action in the explicit form was given.

The aim of this letter is to present AdS$_5$ generalizations of
the two versions of the off-shell minimal Goldstone superfield actions of partially
broken $N=2, d=4$ Poincar\'e supersymmetry: the one with the $N=1$
Goldstone tensor multiplet \cite{BG2,RT,R2} and the one with the
chiral Goldstone $N=1$ supermultiplet \cite{BG1,BG2}.
Instead of dealing with a nonlinear realization of $SU(2,2|1)$
in the standard approach \cite{nonl} like this has been done in \cite{kum},
we prefer to follow the line of refs. \cite{BG2,RT,R2,dik,ik1}. As a first step, we construct
a nonlinear realization of $SU(2,2|1)$ on the set of three
$N=1$ superfields: an improved $N=1$ tensor superfield $L$ and mutually
conjugated chiral superfields $F,\bar F$. This set is subjected to
some nonlinear covariant constraints which leave us with the
single superfield $L$ as the only Goldstone one. Its $SU(2,2|1)$ invariant
action describes $N=1$ L3-brane on AdS$_5$.\footnote{See e.g.  \cite{howe} for the
relevant nomenclature.} The bosonic core of this action is a static-gauge Nambu-Goto
action of L3-brane in AdS$_5$, with one scalar physical field of $L$
being a transverse brane coordinate and another (on-shell) bosonic
degree of freedom being carried out by the notoph field strength.
Then we dualize $L$ into a pair of mutually conjugated chiral $N=1$ superfields and
obtain an analog of the action of ref. \cite{BG1,BG2}. It describes
a scalar super 3-brane on AdS$_5\times S^1$. This action corresponds
to the PBGS option studied in \cite{kum} and in the bosonic sector precisely yields
the $S^5 \rightarrow S^1$ reduction of the scalar part of D3-brane action on
AdS$_5\times S^5$ \cite{ads,mt2}.
\vspace{0.4cm}

\noindent{\bf 2. Goldstone tensor $N=1$ multiplet in a flat background.}
The idea to utilize $N=1$ tensor multiplet as the Goldstone one for describing
the partial breaking of the global $N=2, d=4$ Poincar\'e supersymmetry down to
$N=1$ has been worked out in \cite{BG2,RT,R2}.

One starts with $N=2, d=4$ Poincar\'{e} superalgebra extended by
a real central charge $D$
\be\label{flatalgebra}
 \left\{ Q_{\alpha},{\bar Q}_{\dot\alpha} \right\}=2P_{\alpha\dot\alpha}\;,\;
 \left\{ S_{\alpha},{\bar S}_{\dot\alpha} \right\}=2P_{\alpha\dot\alpha}\;,\;
\left\{ Q_{\alpha},S_{\beta} \right\}=-\ve_{\alpha\beta}D \;,\;
 \left\{ {\bar Q}_{\dot\alpha},{\bar S}_{\dot\beta} \right\}=-\ve_{\dot\alpha\dot\beta}D \;.
\ee
Here $Q_{\alpha},{\bar Q}_{\dot\alpha}$ and $S_{\alpha},{\bar S}_{\dot\alpha}$ are generators
of the unbroken and broken $N=1$ supersymmetries, respectively. These generators
and the 4-translation generator $P_{\alpha\dot\alpha}$ possess the standard
commutation relations with the Lorentz $so(1,3)$ generators
$( M_{\alpha\beta},{\bar M}_{\dot\alpha\dot\beta})$:
\bea
&& i\left[  M_{\alpha\beta},M_{\rho\sigma}\right] =
 \ve_{\alpha\rho}M_{\beta\sigma}+\ve_{\alpha\sigma}M_{\beta\rho}+
         \ve_{\beta\rho}M_{\alpha\sigma}+
           \ve_{\beta\sigma}M_{\alpha\rho} \equiv
       \left( M \right)_{\alpha\beta,\rho\sigma}\;,\nn
&& i\left[  {\bar M}_{\dot\alpha\dot\beta},{\bar M}_{\dot\rho\dot\sigma}\right] =
 \left( {\bar M} \right)_{\dot\alpha\dot\beta,\dot\rho\dot\sigma}\;,\quad
 i\left[  M_{\alpha\beta},P_{\rho\dot\rho}\right] =
\ve_{\alpha\rho}P_{\beta\dot\rho}+\ve_{\beta\rho}P_{\alpha\dot\rho}\;,\nn
&& i\left[  {\bar M}_{\dot\alpha\dot\beta},P_{\rho\dot\rho}\right] =
\ve_{\dot\alpha\dot\rho}P_{\rho\dot\beta}+\ve_{\dot\beta\dot\rho}P_{\rho\dot\alpha}\;,\quad
 i\left[ M_{\alpha\beta}, Q_{\gamma}\right]=\ve_{\alpha\gamma}Q_{\beta}+
  \ve_{\beta\gamma}Q_{\alpha} \equiv \left( Q\right)_{\alpha\beta,\gamma}\;,\nn
&& i\left[ M_{\alpha\beta}, S_{\gamma}\right]= \left( S\right)_{\alpha\beta,\gamma}\;,\quad
  i\left[ {\bar M}_{\dot\alpha\dot\beta}, {\bar Q}_{\dot\gamma}\right]=
  \left( {\bar Q}\right)_{\dot\alpha\dot\beta,\dot\gamma}\;,\quad
 i\left[ {\bar M}_{\dot\alpha\dot\beta}, {\bar S}_{\dot\gamma}\right]=
  \left( {\bar S}\right)_{\dot\alpha\dot\beta,\dot\gamma}\;.
\eea
Then one introduces two $N=1$ superfields: a real one  $L(x,\theta)$
subjected to the constraint
\be
D^2 L = \bar D^2 L =0~, \label{Lconstr}
\ee
and so describing a linear (or tensor) $N=1$ supermultiplet, and a complex
chiral $N=1$ superfield $F, \bar F$,
\be
D_\alpha F = \bar D_{\dot\alpha}\bar F = 0~. \label{chir1}
\ee
Here
\be
D_\alpha = \frac{\partial}{\partial \theta^\alpha} +
i \bar\theta^{\dot\alpha}\partial_{\alpha\dot\alpha}~, \;\;
\bar D_{\dot\alpha} = -\frac{\partial}{\partial \bar\theta^{\dot\alpha}} -
i \theta^{\alpha}\partial_{\alpha\dot\alpha}~, \;\; D^2 = D^\alpha D_\alpha~, \; \bar D^2 =
\bar D_{\dot\alpha}\bar D^{\dot\alpha}~.
\ee
On these $N=1$ superfields one implements \cite{BG2} the following off-shell
representation of the full $N=2$ supersymmetry \p{flatalgebra}:
\be\label{n1tr}
 \delta L = -i\left( \eta^\alpha \theta_\alpha - {\bar\eta}_{\dot\alpha} \tb^{\dot\alpha} \right)+
  \eta^\alpha D_\alpha {\bar F} - {\bar\eta}^{\dot\alpha} \Db_{\dot\alpha} F \;, \;
 \delta F = -\eta^\alpha D_\alpha L \;, \;
\delta {\bar F} = {\bar\eta}^{\dot\alpha} \Db_{\dot\alpha} L \;,
\ee
where $\eta_\alpha, \bar\eta_{\dot\alpha}$ are the infinitesimal transformation
parameters associated with the generators $S_\alpha$, $\bar S_{\dot\alpha}$.
It is a modification of the transformation law of $N=2$ tensor multiplet \cite{tm}
written in terms of its $N=1$ superfield components. This modification is such that
we are in fact facing the {\it Goldstone} $N=2$ tensor multiplet:
the spinor derivatives $D_\alpha L|, \bar D_{\dot\alpha}L|$ are shifted by
$\eta_\alpha, \bar\eta_{\dot\alpha}$ and so are Goldstone fermions for the
partial spontaneous breaking $N=2 \rightarrow N=1$, while $L|$ is shifted by a
constant under the action of the generator $D$ and so is the relevant Goldstone
field ($\vert $ means restriction to the $\theta, \bar\theta$ independent parts).

One can construct the simplest invariant `action' as follows
\be\label{action1f}
S=\frac{1}{4}\int d^4x d^2 \tb F + \frac{1}{4}\int d^4x d^2 \theta {\bar F} \;.
\ee
To make it meaningful one should express the chiral supermultiplet $F,\bar F$
in terms of the Goldstone tensor multiplet $L$ by imposing proper covariant
constraints. These additional constraints were simply guessed in \cite{BG2} and
later re-derived in \cite{RT} from the nilpotency conditions imposed
on the appropriate superfields. They read
\be\label{flatcon}
F=-\frac{D^\alpha L\; D_\alpha L}{2-D^2 {\bar F}} \; \quad
{\bar F}=-\frac{\Db_{\dot\alpha} L\; \Db^{\dot\alpha} L}{2-\Db^2 F}
\ee
and can be easily solved \cite{BG2,RT}
\be\label{sol1flat}
F=- \psi^2+\frac{1}{2}  D^2 \left[
 { {\psi^2{\bar\psi}^2}\over{1+\frac{1}{2}A+\sqrt{1+A+\frac{1}{4}B^2}}}
 \right],
\ee
where
\be
\psi_\alpha \equiv D_{\alpha}L\;, \; {\bar\psi}_{\dot\alpha}\equiv
 {\bar D}_{\dot\alpha}L \;, \;
 A=\frac{1}{2}\left( D^2{\bar\psi}^2+{\bar D}^2\psi^2\right),\;
 B=\frac{1}{2}\left( D^2{\bar\psi}^2-{\bar D}^2\psi^2\right).
\ee
Finally, the action \p{action1f} becomes
\be\label{action2f}
S= -\frac{1}{4}\int d^4xd^2\theta {\bar\psi}^2 -
\frac{1}{4}\int d^4xd^2{\bar\theta} \psi^2 +
\frac{1}{4}\int d^4xd^4\theta
{ {\psi^2}{\bar\psi}^2 \over{1+\frac{1}{2}A+\sqrt{1+A+\frac{1}{4}B^2}}} \;.
\ee
It is a nonlinear extension of the standard $N=1$ tensor multiplet action.
In the bosonic sector it gives rise to the static-gauge Nambu-Goto action for
L3-brane in $d=5$ Minkowski space, with one physical scalar of $L$ being
the transverse brane coordinate and another one represented by the notoph field strength.
After dualizing $L$ into a pair of conjugated chiral and antichiral $N=1$ superfields
(the notoph strength is dualized into a scalar field) the PBGS form
of the worldvolume action of super 3-brane in $d=6$ is reproduced \cite{BG2}.

We would like to point out that the constraints
\p{flatcon} which play the central role in deriving the action \p{action2f}
are intimately related to the 5-dimensional nature of the brane
under consideration. They guarantee 5-dimensional Lorentz covariance.

Indeed, the generator $D$ in \p{flatalgebra} can be treated as the generator
of translations in 5th dimension and the full automorphism algebra
of \p{flatalgebra} can be checked to be $so(1,4)$ (we ignore the $R$-symmetry
$SU(2)$ automorphisms which are explicitly broken in \p{action2f}).
The 5D Lorentz algebra $so(1,4)$ includes,
besides 4D Lorentz generators $M_{\alpha\beta},{\bar M}_{\dot\alpha\dot\beta}$,
an additional 4D vector $K_{\alpha\dot\alpha}$ belonging to the coset $SO(1,4)/SO(1,3)$.
The full set of additional commutation relations is as follows:
\bea\label{addcom}
&&
 i\left[  M_{\alpha\beta},K_{\rho\dot\rho}\right] =
\ve_{\alpha\rho}K_{\beta\dot\rho}+\ve_{\beta\rho}K_{\alpha\dot\rho}\;,\;
i\left[  K_{\alpha\dot\alpha},K_{\beta\dot\beta}\right] =
 -\ve_{\alpha\beta}{\bar M}_{\dot\alpha\dot\beta}-
  \ve_{\dot\alpha\dot\beta}M_{\alpha\beta}\;, \nn
&& i\left[ D, K_{\alpha\dot\alpha}\right] = 2P_{\alpha\dot\alpha}\;,\;
 i\left[  P_{\alpha\dot\alpha},K_{\beta\dot\beta}\right] =
 \ve_{\alpha\beta}\ve_{\dot\alpha\dot\beta}D \;, \nn
&&  i\left[ K_{\alpha\dot\alpha}, Q_{\beta}\right]=-\ve_{\alpha\beta}{\bar S}_{\dot\alpha}\;,\;
i\left[ K_{\alpha\dot\alpha},{\bar S}_{\dot\beta}\right]=
    \ve_{\dot\alpha\dot\beta}Q_{\alpha}\;.
\eea
Now one can check that the following nonlinear transformations
\bea\label{ktrf1}
&& \delta^* L = a_{\alpha\dot\alpha}x^{\alpha\dot\alpha}- a^{\alpha\dot\alpha} \partial_{\alpha\dot\alpha}
 \left( L^2-2 F{\bar F} \right) +i a^{\alpha\dot\alpha}\theta_{\alpha}\Db_{\dot\alpha}F
  -i a^{\alpha\dot\alpha}\tb_{\dot\alpha} D_{\alpha}{\bar F} \; ,\nn
&& \delta^* F=-2 a^{\alpha\dot\alpha}\partial_{\alpha\dot\alpha}\left( FL\right) +
    ia^{\alpha\dot\alpha}\tb_{\dot\alpha}D_{\alpha} L\; , \;
 \delta^* {\bar F}=-2 a^{\alpha\dot\alpha}\partial_{\alpha\dot\alpha}\left( {\bar F}L\right) -
    ia^{\alpha\dot\alpha}\theta_{\alpha}\Db_{\dot\alpha} L
\eea
are just the $SO(1,4)/SO(1,3)$ ones, with $a^{\alpha\dot\alpha}$ being
a transformation parameter related to the additional generator
$K_{\alpha\dot\alpha}$. They have a correct closure on $SO(1,3)$ and
are compatible with the defining constraints \p{Lconstr}, \p{chir1} only provided the
nonlinear constraints \p{flatcon} are imposed. The action \p{action2f} is invariant
under these transformations.
\vspace{0.5cm}

\noindent{\bf 3. AdS$_5$ background.}
Now we wish to generalize the flat superspace construction described in the previous
Section to the case of partial spontaneous breaking of the simplest AdS$_5$
supersymmetry which is $SU(2,2|1)$, that is $N=1$ superconformal group in $d=4$.

The superalgebra $su(2,2|1)$ contains $so(2,4)\times u(1)$
bosonic subalgebra with the generators $\left\{ P_{\alpha\dot\alpha},
M_{\alpha\beta},{\bar M}_{\dot\alpha\dot\beta},
K_{\alpha\dot\alpha},D\right\}$ and $\left\{ J\right\}$ and eight supercharges
$\left\{ Q_\alpha,{\bar Q}_{\dot\alpha},S_\alpha,{\bar S}_{\dot\alpha}\right\}$.
We choose the basis in a such way, that the generators $K_{\alpha\dot\alpha}$ form $so(1,4)$
subalgebra together with the $d=4$ Lorentz generators
$\left\{ M_{\alpha\beta},{\bar M}_{\dot\alpha\dot\beta}\right\}$,
as in the first  line of \p{addcom}. The rest of non-trivial (anti)commutators reads
\bea\label{adsalgebra}
&& i\left[ D, P_{\alpha\dot\alpha}\right] = m P_{\alpha\dot\alpha}\;, \;
 i\left[ D, K_{\alpha\dot\alpha}\right] = 2P_{\alpha\dot\alpha}-m K_{\alpha\dot\alpha}\;,\nn
&&  i\left[  P_{\alpha\dot\alpha},K_{\beta\dot\beta}\right] =
 \ve_{\alpha\beta}\ve_{\dot\alpha\dot\beta}D -\frac{m}{2}\left(
 \ve_{\alpha\beta}{\bar M}_{\dot\alpha\dot\beta}+
  \ve_{\dot\alpha\dot\beta}M_{\alpha\beta}\right), \nn
&& \left\{ Q_{\alpha},S_{\beta} \right\}=-\ve_{\alpha\beta}\left( D+imJ \right)+
m M_{\alpha\beta}\;,\;
 \left\{Q_\alpha, \bar Q_{\dot\alpha}\right\} = 2P_{\alpha\dot\alpha}~, \;
\left\{S_\alpha, \bar S_{\dot\alpha}\right\} = 2P_{\alpha\dot\alpha} - 2m K_{\alpha\dot\alpha}~, \nn
&& i \left[ D, Q_{\alpha}\right]=\frac{m}{2}Q_{\alpha}\;,\;
 i \left[ D, S_{\alpha}\right]=-\frac{m}{2}S_{\alpha}\;,\;
   \left[ J, Q_{\alpha}\right]=\frac{3}{2}Q_{\alpha}\;,\;
    \left[ J, S_{\alpha}\right]=-\frac{3}{2}S_{\alpha}\;,\nn
&& i\left[ K_{\alpha\dot\alpha}, Q_{\beta}\right]=-\ve_{\alpha\beta}{\bar S}_{\dot\alpha}\;,\;
i\left[ K_{\alpha\dot\alpha}, S_{\beta}\right]=\ve_{\alpha\beta}{\bar Q}_{\dot\alpha}\;,\;
 i\left[ P_{\alpha\dot\alpha}, S_{\beta}\right]=m\ve_{\alpha\beta}{\bar Q}_{\dot\alpha}\;.
\eea
This basis is an example of the `AdS basis' of conformal superalgebras
\cite{solvable,solvable1,dik,BIK2} which
perfectly suits their interpretation as the superisometry groups of the appropriate
AdS superspaces. Indeed, the generators $P_{\alpha\dot\alpha}, D, J$ form
a maximal solvable bosonic subgroup in $su(2,2|1)$ and span the coset
$SO(2,4)/SO(1,4)\times U(1)\sim$ AdS$_5\times S^1$. The parameter $m$ has the meaning
of the inverse AdS$_5$ radius, $m= R^{-1}$. In the limit $m=0$ ($R = \infty$) one recovers
from \p{adsalgebra} the $N=1, d=5$ Poincar\'e superalgebra, with $D$
becoming the 5th component of momenta. The generators $J$ and
$K_{\alpha\dot\alpha}, M_{\alpha\beta}, \bar M_{\dot\alpha\dot\beta}$ decouple
and generate outer $u(1)\oplus so(1,4)$ automorphisms.

Our goal is to construct an AdS$_5$ version of the nonlinear
realization \p{n1tr}, \p{flatcon}. The main hints which allowed us
to do this are as follows. Firstly, we assert that this realization
involves some modification of $N=1$ tensor multiplet $L$
and, as before, a pair of mutually conjugated $N=1$ chiral and anti-chiral
superfields $F, {\bar F}$ subjected to some generalization of \p{flatcon}.
Second, in a close analogy with the flat case we require that the following `action'
\be\label{action1}
S\sim \int d^4x d^2{\bar\theta} F + \int d^4x d^2 \theta {\bar F}
\ee
is an invariant of the AdS$_5$ supersymmetry. Since the right-chiral integration measure
$d^4xd^2\bar\theta$ has the $D$ weight $-3m$ and,
with our normalization of $J$, the $U(1)$ charge $-3$, the superfield $F$
should carry the $D$ and $J$ weights equal to $3m$ and $3$ ($\bar F$ has the same $D$ weight and
the $J$ charge equal to $-3$). Third, in the limit $m=0$ our construction
should reproduce the flat case outlined in Sec. 2. At last, it is sufficient
to find the realization of conformal $S$ supersymmetry, since the rest of $SU(2,2|1)$
transformations appears in the closure of these $S$ transformations with themselves and
with those of $N=1$ Poincar\'e supersymmetry.

It turns out that this reasoning almost uniquely fixes the sought transformation
laws and constraints (more details of the derivation are given in
\cite{bik12}). These are
\bea
\delta^* {\bar F} &=& 6im \theta^\alpha\eta_\alpha {\bar F} -
 \Delta x^{\alpha\dot\alpha} \partial_{\alpha\dot\alpha} {\bar F} +
 \Delta\theta^\alpha D_{\alpha} {\bar F} +
  ie^{-2mL}{\bar\eta}^{\dot\alpha}{\bar D}_{\dot\alpha} L \;, \nn
\delta^* F & =& -6im {\bar\theta}_{\dot\alpha}{\bar\eta}^{\dot\alpha} F -
  \Delta x^{\alpha\dot\alpha} \partial_{\alpha\dot\alpha} F -
 \Delta{\bar\theta}^{\dot\alpha} {\bar D}_{\dot\alpha} F +
  ie^{-2mL}{\eta}^\alpha D_\alpha L \;, \nn
\delta^* L& = & -i (\theta^\alpha\eta_\alpha -
 {\bar\theta}_{\dot\alpha}{\bar\eta}^{\dot\alpha}) -
 \Delta x^{\alpha\dot\alpha} \partial_{\alpha\dot\alpha} L+
  \Delta\theta^\alpha D_{\alpha}L -\Delta{\bar\theta}^{\dot\alpha}
   {\bar D}_{\dot\alpha}L \nn
&&  -ie^{2mL}\left[ \eta^\alpha D_\alpha\left( e^{2mL}\bF \right) +
  {\bar\eta}^{\dot\alpha}{\bar D}_{\dot\alpha}
       \left( e^{2mL}F \right)\right], \label{maintr} \\
\frac{1}{m}D^2 e^{-2mL}&=&\frac{1}{m}{\bar D}^2 e^{-2mL}=0\;, \quad D_\alpha F =
\bar D_{\dot\alpha}\bar F =0~, \label{adscon1} \\
 F&=&-{ {e^{-2mL}{D}^{\alpha}L{D}_{\alpha}L }\over
{2-e^{4mL} {D}^2 \bF}} \;, \;
{\bar F}=-{ {e^{-2mL}{\bar D}_{\dot\alpha}L{\bar D}^{\dot\alpha}L }\over
{2-e^{4mL} {\bar D}^2 F}} \;.\label{basiccon}
\eea
Here
\bea
&& \Delta x^{\alpha\dot\alpha}=2im\left( \eta_{\beta}x^{\beta\dot\alpha}
 \theta^{\alpha}+
 {\bar\eta}_{\dot\beta}x^{\alpha\dot\beta}{\bar\theta}^{\dot\alpha}\right)-
 m\left( \theta^2 \eta^\alpha{\bar\theta}^{\dot\alpha} -
 {\bar\theta}^2{\bar\eta}^{\dot\alpha}\theta^{\alpha}\right), \nn
&& \Delta \theta^\alpha=m{\bar\eta}_{\dot\alpha}x^{\alpha\dot\alpha}+
 im\left( \theta^2\eta^\alpha -
 {\bar\theta}_{\dot\alpha}{\bar\eta}^{\dot\alpha}\theta^{\alpha}\right),\;
\Delta {\bar\theta}^{\dot\alpha} =m{\eta}_{\alpha}x^{\alpha\dot\alpha}-
 im\left( {\bar\theta}^2{\bar\eta}^{\dot\alpha} -
 {\theta}^{\alpha}{\eta}_{\alpha}{\bar\theta}^{\dot\alpha}\right)
\eea
are the standard transformations of the $N=1$ superspace coordinates
with respect to the conformal supersymmetry.

In the limit $m=0$ eqs. \p{maintr}, \p{adscon1} and \p{basiccon} go,
respectively, into  \p{n1tr}, \p{Lconstr}, \p{chir1} and \p{flatcon}.
We have checked that, on the surface of the nonlinear constraints \p{basiccon},
the off-shell transformations \p{maintr} are, first, compatible
with the differential constraints \p{adscon1} and, second, produce
the whole $SU(2,2|1)$ symmetry when commuted
among themselves and with $N=1$ Poincar\'e supersymmetry. Had we neglected
the last nonlinear terms in \p{maintr}, we would recover the standard linear $N=1$
superconformal transformation laws of the {\it improved} tensor superfield $e^{-2mL}$
and chiral superfields $F, \bar F$ which close without any need in
the nonlinear constraints \p{basiccon}. It is just due to the presence of these
extra mixed terms the transformations \p{maintr} constitute a realization of $SU(2,2|1)$
as the superisometry group of super AdS$_5$ background and correctly generalize
the flat superspace realization \p{n1tr}. A striking difference between \p{n1tr}
and \p{maintr} lies, however, in the fact that \p{n1tr} close on $N=2$ Poincar\'e
superalgebra before imposing the constraints \p{flatcon}, while \p{maintr} define a
closed supergroup structure only provided the corresponding constraints \p{basiccon}
are imposed from the very beginning. In this sense the situation is similar to
the implementation of the $SO(1,4)$ transformations \p{ktrf1} in the flat case, which
are closed (together with the $SO(1,3)$ ones) only on the surface of \p{flatcon}.
Since in the case of the supergroup $SU(2,2|1)$ these $SO(1,4)$ transformations appear
in the anticommutators of the $Q$ and $S$ supersymmetry generators, it is quite
natural that the constraints \p{basiccon} should enter the game already at the stage
of defining $S$ supersymmetry transformations. It is straightforward to check that \p{basiccon}
by themselves are covariant under the transformations \p{maintr}.

Inspecting \p{maintr}, one can be convinced that this realization just
corresponds to a half-breaking of the $SU(2,2|1)$ supersymmetry: the spinor
derivatives of $L$ are shifted by spinor parameters under the action of
$S$ supersymmetry, thus signaling that the latter is spontaneously broken.
Broken are also $D$ transformations (with $L|$ as the Goldstone field)
and the $SO(1,4)/SO(1,3)$ transformations generated by $K_{\alpha\dot\alpha}$
(with $\partial_{\alpha\dot\alpha} L|$ as the relevant `Goldstone field').

Like their flat counterparts, the constraint \p{basiccon} can be easily solved
\be\label{sol1}
F=-e^{-2mL}{\psi}^2+\frac{1}{2}{ D}^2 \left[
 { {\psi^2{\bar\psi}^2}\over{1+\frac{1}{2}A+\sqrt{1+A+\frac{1}{4}B^2}}}
 \right],
\ee
where
\be
 \psi_\alpha \equiv D_{\alpha}L\;, \quad {\bar\psi}_{\dot\alpha}\equiv
 {\bar D}_{\dot\alpha}L \;, \;
 A=\frac{1}{2}e^{2mL}\left( D^2{\bar\psi}^2+{\bar D}^2\psi^2\right),\;
 B=\frac{1}{2}e^{2mL}\left( D^2{\bar\psi}^2-{\bar D}^2\psi^2\right).
\ee
Finally, the action \p{action1} can be written in the form
\be\label{action2}
S= -\frac{1}{4}\int d^4xd^2\theta e^{-2mL}{\bar\psi}^2 -
\frac{1}{4}\int d^4xd^2{\bar\theta} e^{-2mL}\psi^2 +
\frac{1}{4}\int d^4xd^4\theta
{ {\psi^2}{\bar\psi}^2 \over{1+\frac{1}{2}A+\sqrt{1+A+\frac{1}{4}B^2}}} \;.
\ee
The first two terms in \p{action2} are recognized as
the action of the improved tensor $N=1$ superfield \cite{imptensor}. In the limit
$m=0$ \p{action2} converts into the flat superspace Goldstone
superfield action \p{action2f}.

Defining the bosonic components as
\be
\phi=L|_{\theta=0}\;,\quad \left[D_{\alpha},{\bar D}_{\dot\alpha}\right]
e^{-2mL}|_{\theta=0}=-2mV_{\alpha\dot\alpha} \;,
\ee
where in virtue of \p{basiccon}
\be
\partial_{\alpha\dot\alpha}V^{\alpha\dot\alpha}=0\; \label{notoph} ,
\ee
the bosonic part of \p{action2} proves to be
\be
S_B=\int d^4xe^{-4m\phi}\left[ 1-\sqrt{1+\frac{1}{2}e^{6m\phi}V^2-2e^{2m\phi}
 (\partial \phi)^2 -
 e^{8m\phi} (V^{\alpha\dot\alpha}\partial_{\alpha\dot\alpha}\phi)^2}
 \,\right].\label{bosL3}
\ee
It is a conformally-invariant extension of the static gauge Nambu-Goto action for
L3-brane in $d=5$: the dilaton $\phi$ can be interpreted as a radial brane coordinate, while
$V^{\alpha\dot\alpha}$ is the field strength of notoph which contributes one more scalar
degree of freedom on shell. As is well known, $V^{\alpha\dot\alpha}$ can be dualized in
an off-shell scalar by introducing the constraint \p{notoph} into the action
with a Lagrange scalar multiplier and then eliminating $V^{\alpha\dot\alpha}$
by its algebraic equation of motion. Extending \p{bosL3} as
\be
S_B \quad \Rightarrow \quad S^{dual}_B = S_B + \int d^4x
\lambda \partial_{\alpha\dot\alpha}V^{\alpha\dot\alpha}
\ee
and eliminating $V^{\alpha\dot\alpha}$, after some algebra we get
\be\label{dual2}
S^{dual}_B = \int d^4 x \, |Z|^4\left[1 - \sqrt{-\mbox{det}\left(\eta_{\mu\nu} -
{2\over m^2}\frac{\partial_\mu Z^n\partial_\nu Z^n}{|Z|^4}\right)}\,\right],
\ee
where
\be
Z^1 = r\, \cos \vartheta~, \;\; Z^2 = r\, \sin \vartheta~, \;\; r \equiv e^{-m\phi}~, \;\;\vartheta
\equiv m\,\lambda~, \;\; \eta_{\mu\nu} = \mbox{diag} (+ --- )~.
\ee
The action \p{dual2} is recognized as the $S^5 \rightarrow S^1$ reduction
of the scalar part of the D3-brane action on AdS$_5\times S^5$ \cite{ads},
that is the static-gauge Nambu-Goto action of scalar 3-brane
on AdS$_5\times S^1$.
\vspace{0.4cm}

\noindent{\bf 4. AdS$_5\times S^1$ Goldstone superfield action}. Here
we repeat the above duality transformation at the full superfield level
and obtain in this way an $SU(2,2|1)$ invariant action of Goldstone chiral
$N=1$ superfield which generalizes the action of \cite{BG1,BG2,RT,R2}
and describes a super 3-brane on AdS$_5\times S^1$ superbackground.
We shall be sketchy about details which can be found in \cite{bik12}.
In its basic steps this dualization procedure is similar to the flat superspace
one of \cite{R2}.

We start with the superfield action \p{action2} and relax the constraints for $L$
in \p{adscon1} by adding a Lagrange multiplier term to the superfield Lagrangian
\be\label{relaxaction}
S^{dual}=\frac{1}{4}\int d^4 x d^2 \theta d^2 \tb \left[ -\frac{1}{2m^2} Y\left( \ln Y-1\right) +
\frac{Y^{-4}}{(2m)^4} (D Y)^2 (\Db Y)^2 f+ \frac{Y}{2m}\left( \vfi+\bvfi\right) \right].
\ee
Here
\be\label{deffi}
Y\equiv e^{-2mL}\;,\quad \Db_{\dot\alpha} \vfi= D_\alpha \bvfi=0 \;, \quad
f = \frac{1}{1 +{1\over 2}A + \sqrt{1 + A +{1\over 4}B^2}}~.
\ee
Next we vary the action \p{relaxaction} with respect to $Y$ in order to obtain an
algebraic equation that would allow us to trade $Y$ for $\vfi, \bvfi$. Though the
expression for $Y$ is rather complicated \cite{bik12}, the calculations
are greatly simplified due to the property that only terms bilinear in fermions really contribute to
the dualized action after substitution of this expression back into \p{relaxaction}.
Also, the terms $\sim \bar D^2Y, D^2Y$ can be reabsorbed into a redefinition of chiral
Lagrange multiplier like in the flat case \cite{R2}. Skipping details, the dual action
turns out to be as follows
\bea\label{dualS}
S^{dual}&=&\frac{1}{8}\int d^4 x d^4\theta \left( \frac{e^{m(\vfi+\bvfi)}}{m^2} \right. \nn
&& \left. +\,\frac{\frac{1}{8} (D\vfi)^2(\Db \bvfi)^2}{1-e^{-m(\vfi+\bvfi)}\partial\vfi\partial\bvfi +
\sqrt{ (1-e^{-m(\vfi+\bvfi)}\partial\vfi\partial\bvfi)^2 -e^{-2m(\vfi+\bvfi)}
  (\partial\vfi)^2(\partial\bvfi)^2}} \right).
\eea

This action goes into the flat $N=2 \rightarrow N=1$ chiral Goldstone superfield action
of \cite{BG2,RT,R2} in the limit $m=0$ and is obviously $SU(2,2|1)$ invariant as it was
obtained by dualizing the $SU(2,2|1)$ invariant action \p{action2}.
We do not give the precise form of the $SU(2,2|1)$ transformations of the chiral superfields
$\vfi, \bvfi$ because they look not too illuminating.
However, it is noteworthy that the standard $U(1)$ isometry associated with
the duality transformation, viz. $\delta\vfi =i\alpha$, $\delta\bvfi = -i\alpha$, now
appears in the closure of the $Q$ and $S$ transformations on these Goldstone superfields,
with the imaginary part of $\vfi|$
being the related extra  Goldstone field. It is just the $J$ (or $\gamma_5$) symmetry
of $SU(2,2|1)$, i.e. the duality transformation brings this symmetry from the stability
subgroup into the coset. A similar phenomenon was observed in \cite{ikl}
in the context of the duality between real and complex forms of $N=2$ superconformal mechanics.
The bosonic core of the action \p{dualS}
coincides with \p{dual2} after the identification
\be
\phi= -\frac{1}{2}\left( \vfi+\bvfi \right), \quad \lambda = \frac{i}{2}\left( \vfi - \bvfi \right).
\ee

Thus we conclude that the Goldstone superfield action \p{dualS} describes the option when
the internal $U(1)$ R-symmetry with the generator $J$ is also
broken in addition to the (super)symmetries broken in the action \p{action2}.
The bosonic coset is basically AdS$_5\times S^1 \propto \{x^{\alpha\dot\alpha},\phi\}\times
\{\lambda\}$ and the bosonic part of the action \p{dualS} is just the static-gauge
Nambu-Goto action of a 3-brane on this manifold. This solves the problem of constructing
an invariant Goldstone superfield action for such PBGS option, as it was posed in \cite{kum}.
Note that both the Goldstone superfield actions \p{action2}, \p{dualS} are uniquely restored
from the $SU(2,2|1)$ invariance and do not involve any free parameters, like
their flat superspace counterparts. It is also worth mentioning that the corresponding
Lagrangian densities, once again in tight analogy with the Goldstone superfield
Lagrangians on the Minkowski superbackgrounds, are invariant under $SU(2,2|1)$ only
up to full derivatives and are similar in this respect to WZW or CS Lagrangians.
\vspace{0.4cm}

\noindent{\bf 5. Concluding remarks.} In this note we have presented new nonlinear realizations of
the simplest AdS$_5$ superisometry group $SU(2,2|1)$ in terms of $N=1$ tensor and chiral
Goldstone superfields. We have explicitly given the corresponding minimal Goldstone superfield
actions, for the second option by dualizing the action for the first one, and shown that they
provide a manifestly $N=1$ supersymmetric off-shell superfield form of worldvolume actions
of L3-superbrane on AdS$_5$ and scalar super 3-brane on AdS$_5\times S^1$. The latter is
a truncation of the AdS$_5\times S^5$ D3-action. In the limit of infinite AdS$_5$ radius
these new actions go into their flat superspace counterparts describing the partial breaking of
$N=2, d=4$ supersymmetry down to $N=1$ supersymmetry \cite{BG2,RT,R2,BG1}.

This study can be considered as a first step towards finding out
Goldstone superfield actions for various patterns of partial breaking
of AdS$_5$ supersymmetries. As was already mentioned in \cite{kum},
it is interesting to look for the action corresponding to the half-breaking
of $N=2$ AdS$_5$ supergroup $SU(2,2|2)$ in a supercoset with the
AdS$_5\times S^1$ bosonic part. The basic Goldstone superfield
which we can expect to encounter in this case should be the appropriate generalization
of the $N=2$ Maxwell superfield strength. This action should
be a superconformal version of Dirac-Born-Infeld action describing
the $N=4 \rightarrow N=2$ partial breaking in the flat superspace \cite{bik11,kt}.
In this connection, let us recall that in the flat case there exists one more
$N=2 \rightarrow N=1$ PBGS option associated with the choice of vector $N=1, d=4$
multiplet as the Goldstone one and corresponding to the space-filling
$N=1$ D3-brane \cite{BG3}. No AdS$_5$ analog of this realization can be defined.
The reason is that for achieving $SU(2,2|1)$ invariance one always needs a dilaton
among the worldvolume Goldstone fields and hence within the relevant
$N=1$ Goldstone superfield. In the vector Goldstone $N=2$ supermultiplet there are
two scalar fields and, therefore, the above objection is evaded. An interesting
related problem is to construct PBGS actions for the PP-wave type superbackgrounds
via proper contractions of AdS supersymmetries and their Goldstone
superfield actions.
\vspace{0.4cm}

\noindent{\bf Acknowledgements.} This work was partially supported
by the European Community's Human Potential
Programme under
contract HPRN-CT-2000-00131 Quantum Spacetime,
the INTAS-00-0254 grant and the Iniziativa
Specifica MI12 of the Commissione IV of INFN, grant DFG 436 RUS 113/669, RFBR-DFG grant No. 02-0204002,
as well as RFBR-CNRS grant No. 01-02-22005. E.I. and S.K. thank the
INFN-LNF for warm hospitality during the
course of this work.

\end{document}